\documentclass[10pt,twocolumn,english,american,aps, showpacs]{revtex4}
\usepackage[T1]{fontenc}
\usepackage[latin9]{inputenc}
\usepackage{color}
\usepackage{textcomp}
\usepackage{amsmath}
\usepackage{graphicx}
\usepackage{amssymb}

\makeatletter

\newcommand{\lyxmathsym}[1]{\ifmmode\begingroup\def\b@ld{bold}
  \text{\ifx\math@version\b@ld\bfseries\fi#1}\endgroup\else#1\fi}

\@ifundefined{textcolor}{}
{%
 \definecolor{BLACK}{gray}{0}
 \definecolor{WHITE}{gray}{1}
 \definecolor{RED}{rgb}{1,0,0}
 \definecolor{GREEN}{rgb}{0,1,0}
 \definecolor{BLUE}{rgb}{0,0,1}
 \definecolor{CYAN}{cmyk}{1,0,0,0}
 \definecolor{MAGENTA}{cmyk}{0,1,0,0}
 \definecolor{YELLOW}{cmyk}{0,0,1,0}
 }

\AtBeginDocument{}

\makeatother

\usepackage{babel}

\makeatother

\usepackage{babel}

\begin{document}

\title{Transfer of BECs through discrete breathers in an optical lattice}

\author{H. Hennig$^{\text{1,2,3}}$, J. Dorignac$^{\text{3,4}}$ and D. K.
Campbell$^{3}$}

\affiliation{$^{\text{1}}$Max Planck Institute for Dynamics and Self-Organization,
Göttingen, Germany}

\affiliation{$^{2}$Institute for Nonlinear Dynamics, University of Göttingen,
Germany}

\affiliation{$^{3}$Boston University, Boston, MA 02215, USA}

\affiliation{$^{4}$Laboratoire de Physique Théorique et Astroparticules, CNRS-IN2P3-UMR5207,
Université Montpellier 2, 34095 Montpellier, France}

\date{\today}
\begin{abstract}
We study the stability of a stationary discrete breather (DB) on a
nonlinear trimer in the framework of the discrete nonlinear Schrödinger
equation (DNLS). In previous theoretical investigations of the dynamics
of Bose-Einstein c\textcolor{black}{ondensates in leaking optical
lattices, collisions between a DB and a lattice excitation, e.g.~a
moving breather (MB) or phonon, were studied. These collisions lead
to the transmission of a fraction of the incident (atomic) norm of
the MB through the DB, while the DB can be shifted in the direction
of the incident lattice excitation. Here we show that there exists
a total energy threshold of the trimer, above which the lattice excitation
can trigger the destabilization of the DB and that this is the mechanism
leading to the movement of the DB. Furthermore, we give an analytic
estimate of upper bound to the norm that is transmitted through the
DB. Our analysis explains the results of the earlier numerical studies
and may help to clarify functional operations with BECs in optical
lattices such as blocking and filtering coherent (atomic) beams. }
\end{abstract}

\pacs{67.85.De, 63.20.Pw, 03.75.Lm, 42.65.Tg}

\maketitle

\subsubsection{\textcolor{black}{Introduction}}

\noindent \textcolor{black}{Since the experimental realization of
Bose-Einstein Condensation (BEC) of ultra-cold atoms in optical lattices
(OLs) \cite{Anderson:1995p14092}, many groups of researchers have
achieved an extraordinary level of control over BECs in optical traps
\cite{Morsch:2006p248,Bloch:2008p173,Gericke:2008p15218,Esteve:2008p14646}.
Among other important applications, this control has allowed for the
investigation of analogues of complex solid state phenomena \cite{Eiermann:2004p15104,Carusotto:2000p2469,Cataliotti:2001p105,Greiner:2002p2437,Schumm:2005p14800}.
Technologically, the emerging field of {}``atomtronics'' promises
a new generation of nanoscale devices.}

\textcolor{black}{An important generic feature of nonlinear lattices
is the existence of discrete breathers (DBs), which are spatially
localized, time-periodic, stable (or at least long-lived) excitations
in spatially extended perfectly periodic discrete systems \cite{Campbell:2004p4331,Campbell:2004p7,Flach:2008p4628}.
DBs arise intrinsically from the combination of nonlinearity and the
discreteness of the system. DBs have been observed in a variety of
systems, including Josephson-junction arrays \cite{Trias:2000p3422,Ustinov:2003p4512},
micromechanical systems \cite{Sato:2003p131}, nonlinear waveguide
arrays \cite{Eisenberg:1998p3721,Morandotti:1999p3764}, $\alpha$
helix proteins \cite{Xie:2000p262}, spins in antiferromagnetic solids
\cite{Schwarz:1999p3380,Sato:2004p65} and BEC \cite{Eiermann:2004p15104}.
The existence, stability, and other properties of DBs have been studied
theoretically throughout the last decade \cite{Aubry:1997p4386,Bishop:2003p4241,Dorignac:2004p242,Flach:2008p4628}.
Among other results, it was shown that they act as virtual bottlenecks
which slow down the relaxation processes in generic nonlinear lattices
\cite{Tsironis:1996p4653,Rasmussen:2000p13687,Livi:2006p195,Ng:2009p13686}.}

\textcolor{black}{Many theoretical studies of the dynamics of a BEC
trapped in an OL use the discrete nonlinear Schrödinger equation (DNLS)
to model the system. Several recent studies based on the DNLS have
observed the collision of a stationary DB with a lattice excitation,
e.g.~a moving breather or phonon \cite{Livi:2006p195,Franzosi:2007p174,Ng:2009p13686}.
If the amplitude of the lattice excitation is too small, it is reflected
entirely from the DB. Above a specific threshold amplitude, part of
the incident atomic norm is transmitted through the DB while the DB
is destabilized and shifted by one or few lattice sites towards the
incoming moving breather (MB) }%
\footnote{\textcolor{black}{Indications for the migration of a DB by one or
a few sites towards a lattice excitation can be found as well in a
$\phi^{4}$ nonlinear lattice; see M.~Ivanchenko, O.~Kanakov, V.~Shalfeev
and S.~Flach, Physica D }\textbf{\textcolor{black}{198}}\textcolor{black}{,
120 (2004).}%
}\textcolor{black}{. This transmission process plays a central role
in the occurrence of scale-free atomic avalanches observed for a whole
range of nonlinearities in leaking optical lattices \cite{Ng:2009p13686}.
However, this process has heretofore not been understood analytically.}

\textcolor{black}{In this article we analyze, analytically and numerically,
the collision process of a stationary DB with a lattice excitation.
To this end, we study the nonlinear trimer, i.e.~the DNLS with $M=3$
lattice sites (see e.g.~\cite{Esteve:2008p14646} for an experimental
realization of a similar system). We calculate analytically the threshold
for the destabilization of the DB as well as an upper bound to the
atomic norm that can be transmitted through the DB. The threshold
and the transmission process are described by introducing a `Peierls-Nabarro
energy landscape' which restricts the accessible region of the phase
space for excitations on the trimer. The `local Ansatz', \cite{Flach:1993p209,Flach:1998p285},
an approach applied successfully to studies of DBs on nonlinear lattices,
suggests that the results we find for the trimer will apply to extended
lattices; the agreement of our results with the recent numerical studies
\cite{Ng:2009p13686} confirms this. For the rather large nonlinearity
we shall consider in the sequel, DBs are well localized, and the most
basic and important DBs occupy only three sites. Within the local
Ansatz we consider only the central site and the two neighboring sites
of a DB. This allows us to reduce the high dimensional dynamical problem
involving $M$ sites to the nonlinear trimer. A detailed analysis
of the reduced problem \cite{Flach:1998p285,Flach:1993p209} shows
that (i) the DB corresponds to a trajectory in the phase space of
the full system which is practically embedded on a two-dimensional
toroidal manifold, thus being quasiperiodic in time; (ii) the full
DB can be reproduced accurately within the nonlinear trimer approximation.}

\textcolor{black}{Although we focus here on BECs, our results are
also relevant in a wide range of other contexts in which the DNLS
applies, most prominently coupled nonlinear optical waveguides \cite{Eisenberg:1998p3721,Hennig:1999p6783,Morandotti:1999p3764,Christodoulides:2001p7559,Christodoulides:2003p7511,Kottos:2004p2825}.}

\subsubsection{\textcolor{black}{The Model Hamiltonian}}

\textcolor{black}{The Bose-Hubbard Hamiltonian is arguably the simplest
model that captures the dynamics of a dilute gas of bosonic atoms
in a deep optical lattice, with chemical potential small compared
to the vibrational level spacing (see e.g.~\cite{Bloch:2008p173}
for a review). In the case of weak interatomic interactions (superfluid
limit) and/or a large number of atoms per well (so that the total
number of atoms $N\sim{\cal O}(10^{4}-10^{5})$ is much larger than
the number of wells $M$), a further simplification is available since
the BECs dynamics admits a semi-classical (mean-field) description.
The resulting Hamiltonian describing the dynamics is \begin{equation}
{\cal H}=\sum_{n=1}^{M}[U|\psi_{n}|^{4}+\mu_{n}|\psi_{n}|^{2}]-\frac{J}{2}\sum_{n=1}^{M-1}(\psi_{n}^{*}\psi_{n+1}+c.c.)\,,\label{eq:H0_livi}\end{equation}
 where $n=1,\ldots,M$ is the index of the lattice site, $|\psi_{n}(t)|^{2}\equiv N_{n}(t)$
is the mean number of bosons at site $n$ (also referred to as the
norm $N_{n}(t)$), $U=4\pi\hbar^{2}a_{s}V_{{\rm eff}}/m$ describes
the interaction between two atoms at a single site ($V_{{\rm eff}}$
is the effective mode volume of each site, $m$ is the atomic mass,
and $a_{s}$ is the $s$-wave atomic scattering length), $\mu_{n}$
is the on-site chemical potential, and $J$ is the tunneling amplitude.
The {}``wavefunctions'' $\psi_{n}$ can be used as conjugate variables
with respect to the Hamiltonian ${\cal H}$, leading to a set of canonical
equations \begin{eqnarray}
i\frac{\partial\psi_{n}}{\partial\tau} & = & \frac{\partial{\cal H}}{\partial\psi_{n}^{*}}\nonumber \\
i\frac{\partial\psi_{n}^{*}}{\partial\tau} & = & -\frac{\partial{\cal H}}{\partial\psi_{n}}\,,\label{eq:canon}\end{eqnarray}
 which upon evaluation yields the Discrete Nonlinear Schrödinger Equation\begin{equation}
i\frac{\partial\psi_{n}}{\partial t}=\lambda|\psi_{n}|^{2}\psi_{n}-\frac{1}{2}[\psi_{n-1}+\psi_{n+1}]\,.\label{eq:DNLS}\end{equation}
Here, $\lambda=2U/J$ is the nonlinearity and $t=J\tau$ is the normalized
time. In Eq.~\ref{eq:DNLS} we have set $\mu_{n}=0\,\forall n$. }

\textcolor{black}{The DNLS can be applied to a remarkably large variety
of systems, in particular this mathematical model describes (in the
mean-field limit) the dynamics of a BEC in an OL of size M \cite{Trombettoni:2001p276}.
Experimentally, the tunneling rate $J$ can be adjusted by the intensity
of the standing laser wave field. A powerful tool to modify the on-site
interaction $U$ is via a Feshbach resonance, where the atomic interactions
can be controlled over a large range simply by changing a magnetic
field. A Feshbach resonance involves the coupling of free unbound
atoms to a molecular state in which the atoms are tightly bound. When
the energy levels of the molecular state and the state of free atoms
come closer, the interaction between the free atoms increases. Thus,
the nonlinearity $\lambda$ can be varied experimentally. Here we
will treat the repulsive case explicitly ($\lambda>0$); however,
the attractive case can be obtained via the `staggering' transformation
$\psi_{n}\to(-1^{n})\psi_{n}$ followed by time-reversal $t\to-t$
\cite{Flach:2008p4628}.}

\subsubsection{\textcolor{black}{Equations for the Nonlinear Trimer and Asymptotic
Solutions}}

\textcolor{black}{}%
\begin{figure}
\textcolor{black}{\centering\includegraphics[width=0.8\columnwidth]{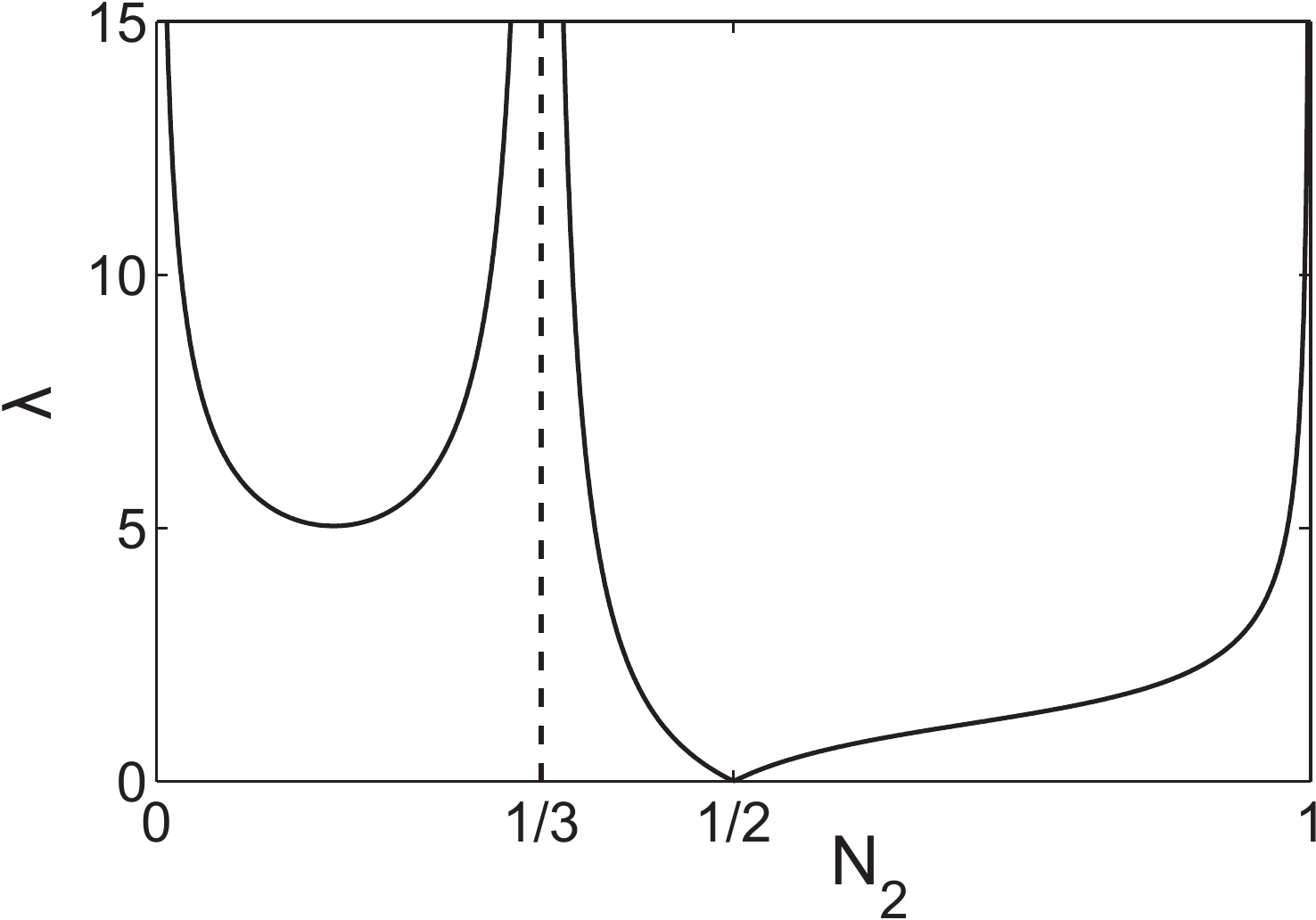}}

\textcolor{black}{\caption{DB solutions for the symmetric case $\psi_{1}=\psi_{3}$ (Eq.~(\ref{eq:lofr2})).
For a nonlinearity $\lambda>5.04$ four symmetric solutions exist.
The solution for $N_{2}>1/2$ is termed a bright breather, while the
solution for $N_{2}\to0$ corresponds to a dark breather (see Eq.~(\ref{eq:psi_breather2})).
The dashed vertical line at $N_{2}=1/3$ marks the asymptote for the
phase-wise and antiphase-wise time-periodic solutions for $\lambda\to\infty$.
\label{fig:lofr}}
}
\end{figure}
\textcolor{black}{To analyze the transfer of norm through a DB (and
related the stability of the DB) during a collision in the nonlinear
trimer ($M=3$) we begin with the equations \begin{eqnarray}
i\partial_{t}\psi_{1} & = & \lambda|\psi_{1}|^{2}\psi_{1}-\frac{1}{2}\psi_{2}\nonumber \\
i\partial_{t}\psi_{2} & = & \lambda|\psi_{2}|^{2}\psi_{2}-\frac{1}{2}(\psi_{1}+\psi_{3})\nonumber \\
i\partial_{t}\psi_{3} & = & \lambda|\psi_{3}|^{2}\psi_{3}-\frac{1}{2}\psi_{2}\,.\label{eq:trimer}\end{eqnarray}
 We normalize the wave functions such that the total atomic population
reads \[
N=\sum_{n=1}^{M}|\psi_{n}|^{2}=1\,.\]
 To find single frequency breather solutions in Eq.~(\ref{eq:trimer})
for the symmetric case $\psi_{1}=\psi_{3}$, we assume \begin{equation}
\psi_{n}(t)=A_{n}e^{iwt}\;,\label{eq:ansatz}\end{equation}
 with amplitudes $A_{n}$ and frequency $w$. This Ansatz, together
with the conservation of particle number, leads to the set of equations\begin{eqnarray}
-wA_{1} & = & \lambda A_{1}^{3}-\frac{1}{2}A_{2}\nonumber \\
-wA_{2} & = & \lambda A_{2}^{3}-A_{1}\nonumber \\
1 & = & 2A_{1}^{2}+A_{2}^{2}\;.\label{eq:ansatz2}\end{eqnarray}
}

\noindent \textcolor{black}{Let us first calculate the relation between
the (atomic) norm $N_{2}=A_{2}^{2}$ at the central site and the nonlinearity
$\lambda$. From Eq.~(\ref{eq:ansatz2}), we find \begin{equation}
\lambda(N_{2})=\pm\frac{\sqrt{2}(2N_{2}-1)}{\sqrt{N_{2}(1-N_{2})}(3N_{2}-1)}\,.\label{eq:lofr2}\end{equation}
We have four solutions above the bifurcation point at $\lambda\approx5.04$
and two solutions for $0\le\lambda<5.04$ (see Fig.~\ref{fig:lofr}).
To gain further insight into the nature of the symmetric solutions
in the trimer, we will revisit Eq.~(\ref{eq:ansatz2}), which we
convert into a quartic equation\begin{equation}
x^{4}+\frac{\lambda}{\sqrt{2}}x^{3}-\sqrt{2}\lambda x-1=0\,,\label{eq:quartic2}\end{equation}
 where\begin{equation}
A_{2}=\cos(\arctan(x))=\frac{\text{sign}(x)}{\sqrt{1+x^{2}}}\,.\label{eq:A2}\end{equation}
 Expansion of the exact real solutions of Eq.~(\ref{eq:quartic2})
in $\lambda$ for the limiting case $\lambda\to0$ gives\begin{eqnarray}
x_{1} & = & 1+\frac{\lambda}{4\sqrt{2}}-\frac{5}{64}\lambda^{2}+{\cal O}(\lambda^{3})\nonumber \\
x_{2} & = & -1+\frac{\lambda}{4\sqrt{2}}+\frac{5}{64}\lambda^{2}+{\cal O}(\lambda^{3})\:.\label{eq:limit0_2}\end{eqnarray}
 At $\lambda\!=\!0$ the solution $\vec{\psi}(t)\!=\!(\psi_{1}(t),\psi_{2}(t),\psi_{3}(t))$
of Eq.~(\ref{eq:ansatz2}) at time $t\!=\!0$ reads $\vec{\psi}_{(x1,x2)}(0)\!=\!(1/2,\pm1/\sqrt{2},1/2)$.
The antisymmetric breather configuration $\vec{\psi}(0)=(-1/\sqrt{2},0,1/\sqrt{2})$
is not included in our Ansatz, as we restrict ourselves to symmetric
solutions.}

\noindent \textcolor{black}{Expansion around the limit $\lambda\to\infty$
leads to four real solutions\begin{eqnarray}
x_{3} & = & -\frac{1}{\sqrt{2}}\frac{1}{\lambda}-\frac{1}{4\sqrt{2}}\frac{1}{\lambda^{3}}{\cal +O}(\lambda^{-5})\nonumber \\
x_{4} & = & -\frac{1}{\sqrt{2}}\lambda+2\sqrt{2}\frac{1}{\lambda}+14\sqrt{2}\frac{1}{\lambda^{3}}{\cal +O}(\lambda^{-5})\nonumber \\
x_{5} & = & -\sqrt{2}-\frac{3}{2\sqrt{2}}\frac{1}{\lambda}-\frac{69}{16\sqrt{2}}\frac{1}{\lambda^{2}}{\cal +O}(\lambda^{-3})\nonumber \\
x_{6} & = & \sqrt{2}-\frac{3}{2\sqrt{2}}\frac{1}{\lambda}+\frac{69}{16\sqrt{2}}\frac{1}{\lambda^{2}}{\cal +O}(\lambda^{-3})\;.\label{eq:limitinf2}\end{eqnarray}
 For infinite $\lambda$ the solutions of Eq.~(\ref{eq:ansatz2})
at time $t=0$ are \begin{eqnarray}
\vec{\psi}_{(x3)}(0) & = & (0,1,0)\nonumber \\
\vec{\psi}_{(x4)}(0) & = & (\sqrt{1/2},0,\sqrt{1/2})\nonumber \\
\vec{\psi}_{(x5,x6)}(0) & = & (1/\sqrt{3},\mp1/\sqrt{3},1/\sqrt{3})\;,\label{eq:psi_breather2}\end{eqnarray}
 where the solution $\vec{\psi}_{(x3)}$ is called a bright breather,
$\vec{\psi}_{(x4)}$ is named a dark breather (due to lack of norm
at the central site) and $\vec{\psi}_{(x5,x6)}$ are phase-wise and
antiphase-wise time-periodic solutions.}

\subsubsection{\textcolor{black}{The Peierls-Nabarro energy landscape}}

\textcolor{black}{Having found the symmetric DB solutions, we next
focus on the transfer of norm through a bright breather, where the
stability of the breather will play a crucial role. We start by introducing
the concept of a `Peierls-Nabarro (PN) energy landscape'.  It is related
to the PN potential, which reflects the fact that discreteness breaks
the continuous translational invariance of a continuum model \cite{Kivshar:1993p91,Rumpf:2004p191}.
The amplitude of the PN potential may be seen as the minimum barrier
which must be overcome to translate an object by one lattice site.
Regarding DBs, the Peierls-Nabarro barrier is given by the energy
difference $|E_{c}-E_{b}|$, where $E_{c}$ and $E_{b}$ are the energies
of a DB centered at a lattice site and between two lattice sites.}

\textcolor{black}{}%
\begin{figure}[t]
\textcolor{black}{\centering\includegraphics[width=1\columnwidth]{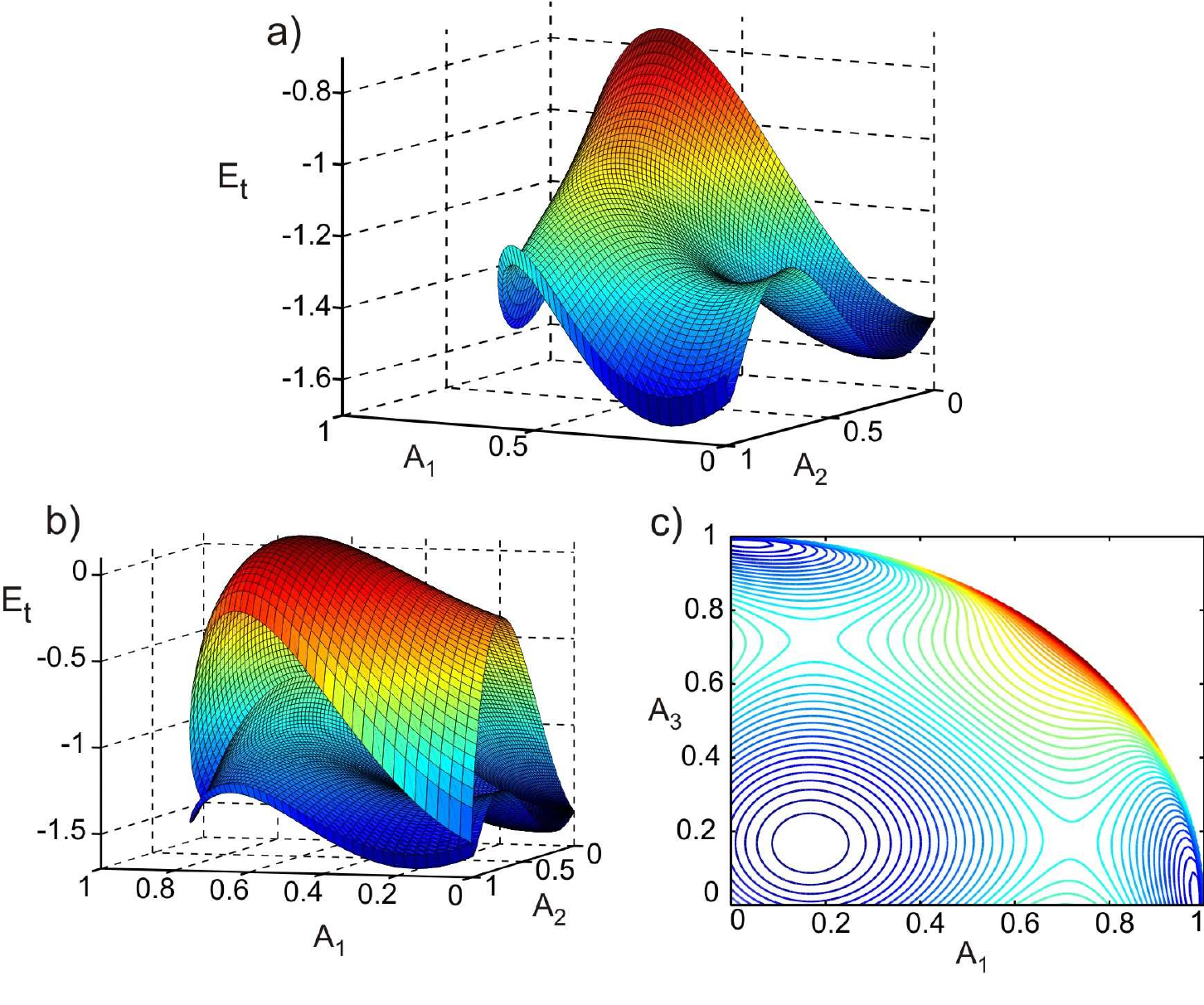}}

\textcolor{black}{\caption{\textbf{(a)} The lower part of the PN energy landscape exhibits three
minima separated by saddle points. \textbf{(b)} The phase space of
the trimer is restricted to lie between the two parts of the PN shell,
which consists of the lower and upper part of the PN landscape, which
are shown in this panel. \textbf{(c)} Contour plot of (a),the lower
part of the PN energy landscape. The three minima and saddle points
are clearly visible. The minimum at $A_{1}=A_{3}=0.17$ in (c) corresponds
to the bright breather. The figure is plotted for $\lambda=3$.\label{fig:PN}}
}
\end{figure}

\textcolor{black}{We define the Peierls-Nabarro energy landscape as
follows: for a given configuration of amplitudes, $A_{n}$, the PN
energy landscape is obtained by extremizing $H$ with respect to the
phase differences $\delta\phi{}_{ij}=\phi_{i}-\phi_{j}$:\begin{equation}
H_{\text{PN}}^{l}=\min_{\delta\phi{}_{ij}}(-H)\quad;\quad H_{pn}^{u}=\max_{\delta\phi{}_{ij}}(-H)\,,\label{eq:PNdef}\end{equation}
 where $\psi_{n}=A_{n}\exp(i\phi_{n})$ and $H_{\text{PN}}^{l}$ and
$H_{\text{PN}}^{u}$ are the lower and upper part of the PN landscape.
As we will see later, the bright breather solution $\vec{\psi}_{(x3)}$
is located at an extremum of $H_{\text{PN}}^{l}$. The minus sign
in the definition (\ref{eq:PNdef}) was added for convenience to assure
that the bright breather is found in a minimum (and not in a maximum)
of the lower PN landscape. The phase differences extremizing the Hamiltonian\begin{eqnarray}
H & = & \frac{\lambda}{2}(A_{1}^{4}+A_{2}^{4}+A_{3}^{4})\nonumber \\
 &  & -(A_{1}A_{2}\cos(\phi_{1}\!-\!\phi_{2})+A_{2}A_{3}\cos(\phi_{2}\!-\!\phi_{3}))\label{eq:Hamiltonian}\end{eqnarray}
 are $\delta\phi_{12}=\delta\phi_{23}\in\{0,\pi\}$. Hence, the upper
and the lower PN energy landscapes read\begin{equation}
H_{\text{PN}}^{u}=-\frac{\lambda}{2}(A_{1}^{4}+A_{2}^{4}+A_{3}^{4})+(A_{1}+A_{3})A_{2}\label{eq:HPNup}\end{equation}
}

\textcolor{black}{and\begin{equation}
H_{\text{PN}}^{l}=-\frac{\lambda}{2}(A_{1}^{4}+A_{2}^{4}+A_{3}^{4})-(A_{1}+A_{3})A_{2}\,.\label{eq:HPN}\end{equation}
}%
\begin{figure*}
\textcolor{black}{\centering\includegraphics[width=1.6\columnwidth]{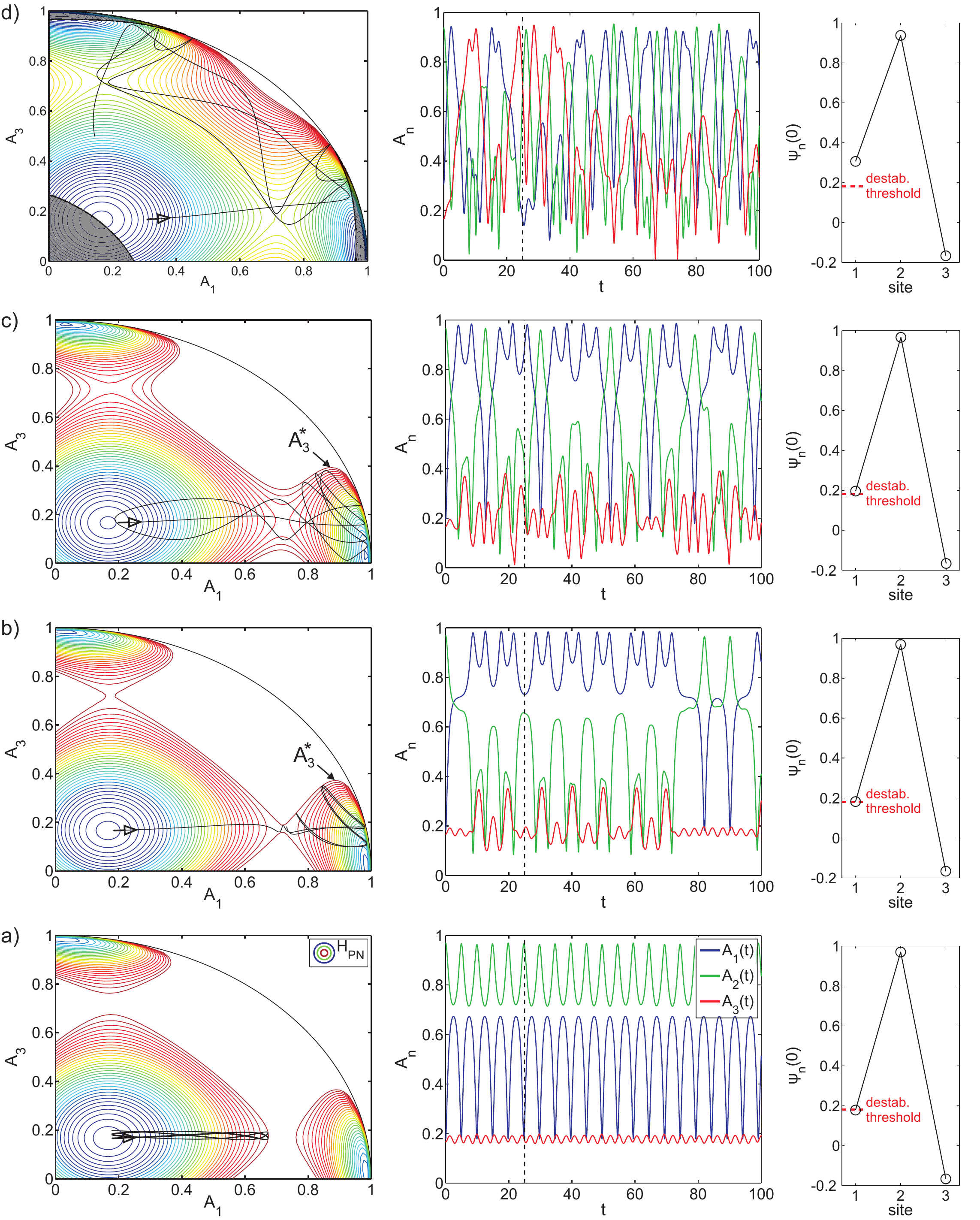}}

\textcolor{black}{\caption{Dynamics on the PN landscape for increasing total energy of the trimer.
\textbf{(a)} \textbf{(left)} A contour plot of the lower PN energy
landscape $H_{\text{PN}}^{l}$ is shown for total energy below the
rim ($E_{t}=-1.32<E_{\text{thrs}}=-1.311$). A projection of the orbit
onto the $A_{1}$--$A_{3}$ plane is over-plotted (black curve). \textbf{(middle)}
The corresponding amplitudes $A_{i}(t)$ indicate that the maximum
amplitude remains at the central site. The dashed vertical line marks
the time interval $[0,25]$ for which the orbits in the left picture
are plotted. \textbf{(right)} A sketch of the initial condition shows
that the excitation at site $1$ is slightly below threshold.\textbf{
(b)} Destabilization of the DB for $E_{t}=-1.310>E_{\text{thrs}}$
just above the rim. We see that the rim of the PN landscape clearly
restricts the dynamics and governs the destabilization process of
the DB. \textbf{(c)} For higher total energy $E_{t}=-1.28$ the bottleneck
at the rim widens and the maximum norm transmitted to site $3$ is
increased. \textbf{(d)} For even higher total energy $E_{t}=-1.04$
the orbit explores large parts of the phase space\textcolor{black}{{}
and visits all three sites. The grey shaded areas are forbidden by
the upper PN landscape $H_{\text{PN}}^{u}$. In all cases $\lambda=3,$
$\delta_{\phi}=\pi$. For other v}alues of $\delta_{\phi}$ the same
qualitative behavior is found. \label{fig:bubble}}
}
\end{figure*}

\textcolor{black}{In Fig.~\ref{fig:PN} the PN landscape is visualized
for $\lambda=3$. The PN `shell', consisting of the upper and lower
landscapes, bounds the phase space of the trimer. Since the DB whose
properties we are studying corresponds to a minimum on $H_{\text{PN}}^{l}$,
we shall focus on this landscape. As shown in Fig.~\ref{fig:PN}
(c), the projection onto the $A_{1}-A_{3}$ plane exhibits three minima
which are separated by saddle points (called `rims' in the following).
For $\lambda\to\infty$ the saddle points are located at $A_{1}=A_{2}=\sqrt{1/2}$
(which in the following will be the saddle point of interest) and
$A_{2}=A_{3}=\sqrt{1/2}$. The energy threshold $E_{\text{thrs}}$
at the rim (obtained from Eq.~(\ref{eq:HPN})) reads\begin{equation}
E_{\text{thrs}}=-\frac{\lambda}{4}-\frac{1}{2}-\frac{1}{4\lambda}+\frac{1}{4\lambda^{2}}-\frac{1}{4\lambda^{3}}+\frac{9}{16\lambda^{4}}+{\cal O}(\lambda^{-5})\,.\label{eq:Erim}\end{equation}
}

\textcolor{black}{The following investigation holds for an effective
nonlinearity in a range around $\Lambda=\lambda/M\simeq1$ }%
\footnote{\textcolor{black}{For nonlinearities $\Lambda\ll1$ no such saddle
points in the PN landscape are found.}%
}\textcolor{black}{, which is in the critical regime where scale-free
avalanches of BECs were found in \cite{Ng:2009p13686}.}

\subsubsection{\textcolor{black}{The Threshold for transfer of norm}}

\textcolor{black}{To study the influence of the PN landscape on the
stability and the transfer of atoms through the DB, we first consider
the fixed point corresponding to the bright breather. The initial
amplitudes $A_{i}^{b}$ are obtained by inverting Eq.~(\ref{eq:lofr2})
for $N_{2}>1/2$; hence an initial condition for the bright breather
reads $\vec{\psi}^{b}(0)\!=\!(-A_{1}^{b},A_{2}^{b},-A_{1}^{b})$.
Then perturbations are added to site $1$. In dynamical systems terminology,
the phase space of the trimer is `mixed', consisting of regular islands
surrounded by the chaotic sea \cite{Mossmann:2006p84,Ng:2009p13686}.
DBs are located inside the regular islands of the phase space, provided
that their frequency (and multiples of their frequency) lie outside
the phonon spectrum \cite{Flach:1993p209,Flach:1998p285}. If a perturbation
is large enough, it can push the orbit out of the regular island into
the chaotic sea, destabilizing the DB.}

\textcolor{black}{We now use the following initial condition:\begin{equation}
\vec{\psi}(0)=(-(A_{1}^{b}+\delta_{A})e^{i\delta_{\phi}},A_{2},-A_{1}^{b})\,,\label{eq:psi_initial}\end{equation}
 where $A_{2}=(1-|\psi_{1}|^{2}-|\psi_{3}|^{2})^{1/2}$ ensures total
norm $N=1$. Compared to the bright breather, we have added an amplitude
$\delta_{A}$ to site $1$ and the phase $\phi_{1}$ is rotated by
$\delta_{\phi}$. The initial condition (\ref{eq:psi_initial}) is
visualized in Fig.~\ref{fig:bubble}(right panels), where we have
fixed $\delta_{\phi}=\pi$ and increased $\delta_{A}$ in (a-d). Note
that although the phase rotation does not alter the norms $|\psi_{i}|^{2}$,
it drastically changes the total energy of the trimer which we define
as $E_{t}=-H$ (see Eq.~(\ref{eq:Hamiltonian})).}

\textcolor{black}{}%
\begin{figure}
\textcolor{black}{\centering\includegraphics[width=0.7\columnwidth]{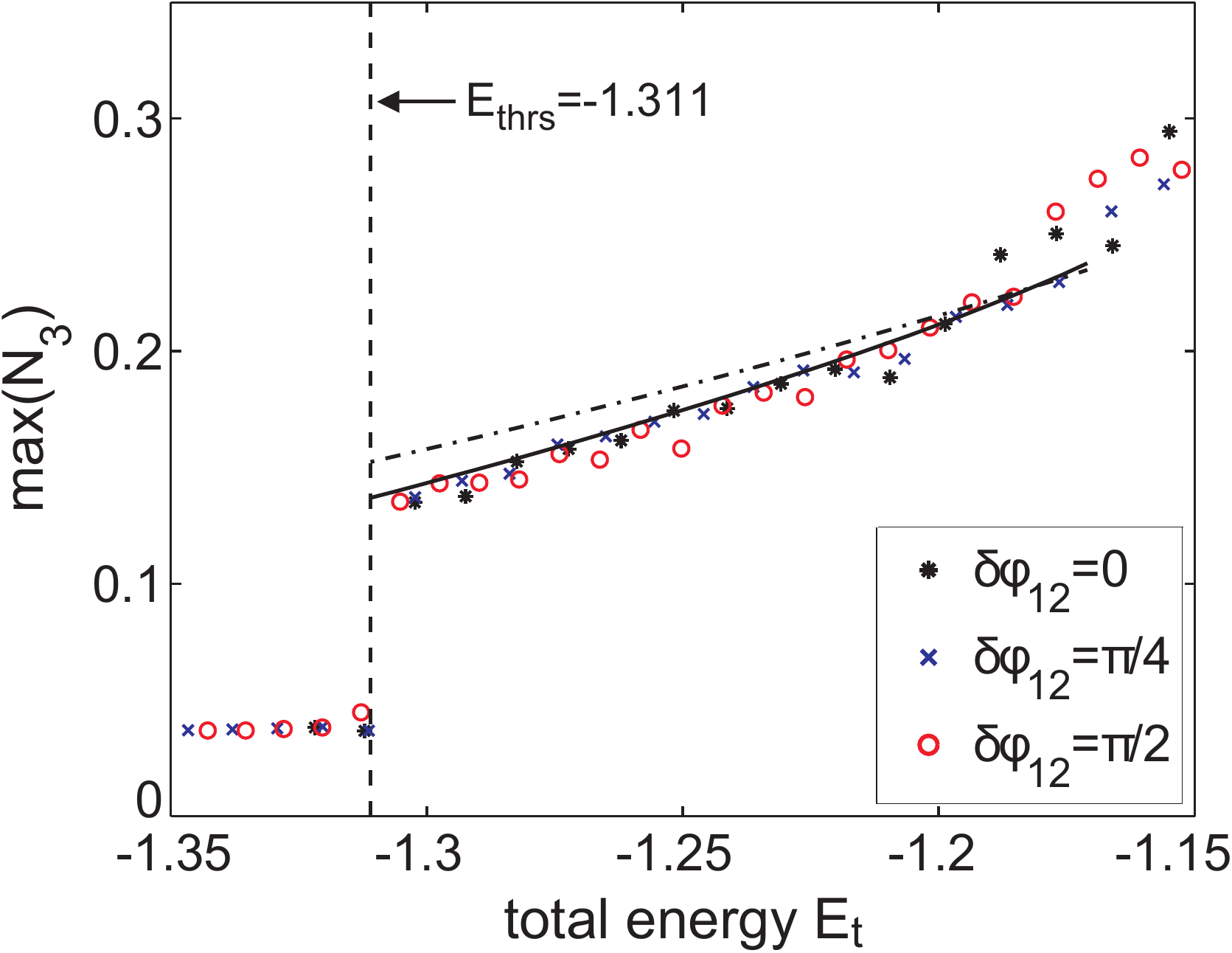}}

\textcolor{black}{\caption{Norm transfer through a bright breather. The maximum (atomic) norm
at site $3$ detected after the collision of the DB with a lattice
excitation is shown as a function of the total energy of the trimer.
The discrete symbols indicate three different values of the initial
phases. For $E_{t}<E_{\text{thrs}}$ the DB is stable and practically
no transfer of norm takes place on short timescales. For $E_{t}>E_{\text{thrs}}$
we observe instability of the DB centered at site $2$: The breather
migrates to site $1$ and norm is transferred to site $3$. An upper
bound to $\max(N_{3}(t))$ is calculated from the PN landscape, both
analytically (dotted dashe\textcolor{black}{d line, cf.~Eq.~(\ref{eq:A3bubble}))
and n}umerically (solid line). The analytical calculation is performed
in the limit for large $\lambda$ and therefore deviates slightly
from the exact numerical result. We used $\delta\phi_{23}=\pi$ which
is the typical case observed in \cite{Ng:2009p13686} for DBs in an
extended leaking optical lattice and $\lambda=3$.\label{fig:ttet}}
}
\end{figure}
\textcolor{black}{In Fig.~\ref{fig:bubble}(left) we show the dynamics
for increasing total energy $E_{t}$, where the arrows on the orbits
(black curves) mark the direction of time. For $E_{t}<E_{\text{thrs}}$
the areas in phase space are disconnected, leading to sub-threshold
dynamics depicted in Fig.~\ref{fig:bubble}a.}

\textcolor{black}{In contrast, for $E_{t}>E_{\text{thrs}}$ the orbit
is allowed to pass the rim such that the majority of the norm migrates
from site $2$ to site $1$ (Fig.~\ref{fig:bubble}b), while norm
is transferred to site $3$. The larger $E_{t}$ (Fig.~\ref{fig:bubble}c),
the larger is the size of the {}``bubble'' (by the term bubble we
denote the accessible region of the PN landscape for $A_{1}>1/\sqrt{2}$).
Hence, as we see from Fig.~\ref{fig:bubble}b-c(left), an upper limit
to the norm that can possibly be transmitted through the DB can be
read from the maximum value of $A_{3}$ inside the bubble\begin{equation}
A_{3}^{\ast}=\max_{A_{1}>1/\sqrt{2}}A_{3}\label{eq:Astar}\end{equation}
for fixed total energy }\foreignlanguage{english}{\textcolor{black}{$E_{t}$}}\textcolor{black}{.}

\textcolor{black}{Finally, for even larger $E_{t}$, the orbit visits
large parts of the phase space and large amplitudes $A_{i}(t)$ are
found at all three sites, as depicted in Fig.~\ref{fig:bubble}d,
where most of the orbit resides in the chaotic regime (Fig.~\ref{fig:bubble}d).
In this case, no controlled shift of the DB from site $2$ to $1$
is observed and the DB becomes dynamically unstable. These dynamical
instabilities can be associated with a (partial) depletion of the
BEC \cite{Trimborn:2009p14451,Trimborn:2008p17866}; a detailed study
of these effects is beyond the scope of our present investigation.}

\textcolor{black}{The upper bound $A_{3}^{\ast}$ can be calculated
analytically noting that for the projection onto the $A_{1}-A_{3}$
plane, condition $dA_{3}/dA_{1}=0$ holds. Implicit derivation of
Eq.~\ref{eq:HPN} leads to \begin{eqnarray}
0 & = & 1-2A_{1}^{2}-A_{1}A_{3}-A_{3}^{2}\nonumber \\
 &  & +2\lambda A_{1}\sqrt{1-A_{1}^{2}-A_{3}^{2}}(-1+2A_{1}^{2}+A_{3}^{2})\,,\label{eq:implicit}\end{eqnarray}
that determines the maximum value of $A_{3}$ in the {}``bubble''
of the PN landscape as\begin{eqnarray}
A_{3}^{\ast}(\delta,\lambda) & = & \frac{1}{\sqrt{2}}-\frac{\sqrt{(1-2\delta)}}{2\sqrt{\lambda}}-\frac{(1-2\delta)}{4\sqrt{2}\lambda}\nonumber \\
 &  & +\frac{1+2(\delta-\delta^{2})}{4\sqrt{1-2\delta}\lambda^{3/2}}-\frac{25-20(\delta-\delta^{2})}{64\lambda^{2}}\nonumber \\
 &  & +\frac{11/8-3\delta-\delta^{2}+7(\delta^{3}-7\delta^{4})}{2(1-2\delta)^{3/2}\lambda^{5/2}}\nonumber \\
 &  & +{\cal O}(\lambda^{-3})\,,\label{eq:A3bubble}\end{eqnarray}
where $\delta=E_{t}-E_{thrs}>0$ is the energy relative to the destabilization
threshold. Without the $\lambda^{-5/2}$ term the exact value is underestimated.
If we truncate the expression after the $\lambda^{-3/2}$ term, the
deviation from the exact result roughly doubles compared to what is
shown in Fig.~\ref{fig:ttet}.}

\textcolor{black}{}%
\begin{figure*}
\textcolor{black}{\includegraphics[width=1.44\columnwidth]{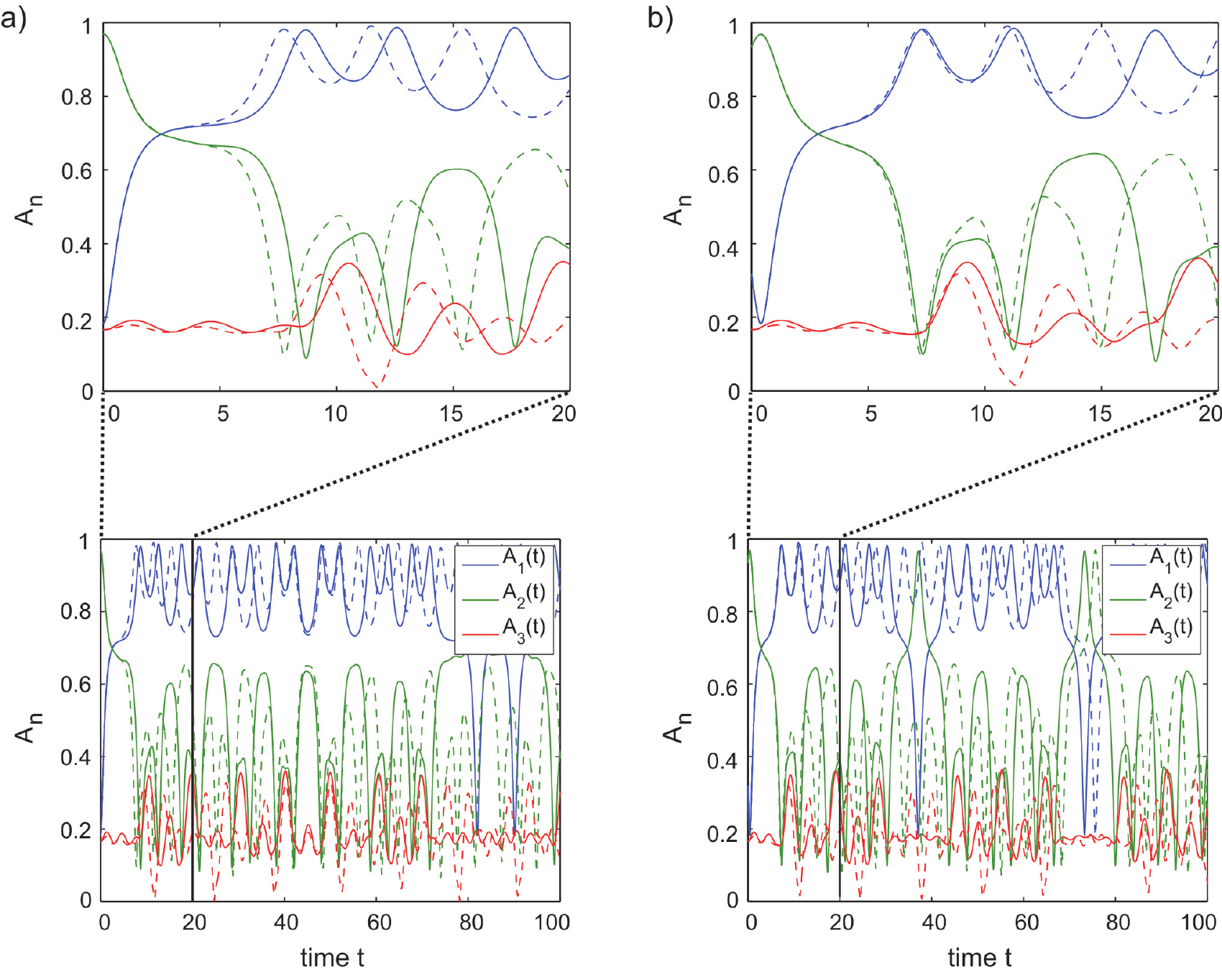}}

\textcolor{black}{\caption{The de\textcolor{black}{stabilization of the DB in the trimer (solid
line) is compared to the trimer with a third linear site (dashed line).
On the timescale, where the destabilization takes place, the dynamics
is qualitatively very similar (see enlargements in upper pictures).
Total energy is $E_{t}=-1.31>E_{\text{thrs}}$ and $\lambda=3$. }\textbf{\textcolor{black}{(a)}}\textcolor{black}{{}
The initial condition (\ref{eq:psi_initial}) is given by $\delta A=0.016$,
$\delta_{\phi}=\pi$, which is the same as in Fig.~\ref{fig:bubble}b.
}\textbf{\textcolor{black}{(b)}}\textcolor{black}{{} The} initial condition
is determined by the parameters $\delta A=0.152$, $\delta_{\phi}=\pi/2$.
\label{fig:linear_site_three}}
}
\end{figure*}
\textcolor{black}{How general is the transfer mechanism that we describe?
In Fig.~\ref{fig:ttet}, the maximum norm $N_{3}$ that is found
at site $3$ after the transfer of atoms through the DB is shown as
a function of the total energy for three initial phase differences
$\delta\phi_{12}=\pi-\delta_{\phi}=0,\,\pi/4,\,\pi/2$. The dashed
vertical line at $E_{t}=E_{\text{thrs}}=-1.311$ marks the total energy
at the rim and is identified with the destabilization threshold of
the DB. Evidently, the transfer mechanism found does not depend on
parameters $\delta_{A}$ and $\delta_{\phi}$ individually, but rather
on the total energy which determines the accessible region of the
PN landscape. Moreover, the transfer mechanism itself appears nearly
independent of the choice of the initial phase difference $\delta\phi_{12}$.
The maximum norm detected at site $3$ is closely below the upper
bound $N_{3}^{\text{\ensuremath{\ast}}}(\delta)=|A_{3}^{\text{\ensuremath{\ast}}}(\delta)|^{2}$
given by Eq.~\ref{eq:A3bubble} which holds for $E_{t}\lesssim-1.2$.
For increasing $E_{t}\gtrsim-1.2$ the fluctuations of $\max(N_{3})$
for different orbits with similar total energy become larger, because
the orbits explore a larger part of the phase space. As a consequence,
$\max_{t<T}(N_{3}(t))$ depends on the chosen time interval $[0,T]$
in which the maximum norm is detected (In all cases we set $T=100$).
In contrast, for lower total energy where a nontrivial upper bound
$N_{3}^{\ast}(\delta)$ of the transferred norm holds, $\max_{t<T}(N_{3}(t))$
does barely depend on $T$ already after very few oscillations of
$A_{3}(t)$ (cf.~Fig.~\ref{fig:bubble}b-c(middle)).}

\subsubsection{\textcolor{black}{Connection to PN barrier}}

\textcolor{black}{In order to gain further insight into the relation
between the rim of the PN energy landscape and the PN barrier, let
us consider a nonlinear trimer where we omit the nonlinear on-site
interaction term at site $3$. The equations of motion read\begin{eqnarray}
i\partial_{t}\psi_{1} & = & \lambda|\psi_{1}|^{2}\psi_{1}-\frac{1}{2}\psi_{2}\nonumber \\
i\partial_{t}\psi_{2} & = & \lambda|\psi_{2}|^{2}\psi_{2}-\frac{1}{2}(\psi_{1}+\psi_{3})\nonumber \\
i\partial_{t}\psi_{3} & = & -\frac{1}{2}\psi_{2}\,.\label{eq:trimer_lin}\end{eqnarray}
 Using the initial condition (\ref{eq:psi_initial}) we find that
on the timescale where the destabilization process of the DB centered
at site $2$ takes place, the dynamics is not changed substantially
compared to the results for the nonlinear trimer (see Fig.~\ref{fig:linear_site_three}).
Hence, in order to describe the destabilization of the DB (and the
basic mechanism of the norm transfer through the DB), it is sufficient
to consider only two nonlinear sites with a third linear site attached.
This is a strong indication that the destabilization threshold during
the collision of the two objects can actually be linked to the Peierls-Nabarro
barrier of a }\textit{\textcolor{black}{single}}\textcolor{black}{{}
DB.}

\subsubsection{\textcolor{black}{Possible Applications}}

\textcolor{black}{To start with, we would like to comment on the validity
of the DNLS (\ref{eq:DNLS}) to describe actual experiments of BECs
in OLs. Experimental realizations have been performed for values of
$\lambda=2U/J$ in the range $10^{\lyxmathsym{\textminus}5}\lyxmathsym{\textendash}10^{\lyxmathsym{\textminus}3}$,
while the number of atoms is typically $N\sim{\cal O}(10^{4}-10^{5})$.
These estimations lead to experimentally feasible parameters $\Lambda=\lambda/M\lesssim1$
for which the DNLS is a good approximation. For example, the experiment
of \cite{Cataliotti:2001p105} shows that the the BEC dynamics in
an OL with parameters $N\approx2\times10^{5}$, $J=0.14E_{R}$, $2UN\approx12E_{R}$
(where $E_{R}=\hbar^{2}k_{L}^{2}/(2m)$ is the recoil energy and $k_{L}$
is the laser mode which traps the atoms), and $M=200$ wells is described
very well by the DNLS with effective nonlinearity $\Lambda\approx0.5$.
To estimate the minimum duration of an experiment with BECs probing
the destabilization process, we rewrite our dimensionless time $t=J\tau$
in terms of real time $\tau$. Typical values for $J$ are taken from
\cite{Gati:2006p15593}: The Josephson energy $E_{J}/k_{B}=378\,\text{nK}$
leads to the tunneling rate $J/\hbar=E_{j}(N\hbar)^{-1}=16.5\,\text{Hz}$
(for $N=3000$ atoms). Hence, with these parameters, $t=10$ (which
is a typical timescale after which the destabilization process took
place) relates to $\tau\approx0.6\,\text{sec}$.}

\textcolor{black}{The thresholded transfer of norm through a bright
breather that we analyzed may lead to interesting applications for
blocking and filtering atom beams. It could be a powerful tool for
controlling the transmission of matter waves in interferometry and
quantum information processes \cite{Vicencio:2007p207}.}

\textcolor{black}{In a similar way, our findings can be related to
the field of optics, as the DNLS is capable of describing wave motion
in nonlinear optical waveguide arrays. Discrete breathers in such
two-dimensional networks were investigated in the past years both
theoretically and experimentally \cite{Kivshar:1993p3978,Eisenberg:1998p3721,Hennig:1999p6783,Morandotti:1999p3764,Christodoulides:2001p7559,Christodoulides:2003p7511,Kottos:2004p2825,Meier:2004p7976}
and can exhibit a rich variety of functional operations such as blocking,
routing or logic functions \cite{Christodoulides:2001p7559,Christodoulides:2003p7511}.
Experimental evidence of the destabilization process of the stationary
DB should be observable in nonlinear waveguide arrays and might lead
to functional operations such as filtering optical beams.}

\textcolor{black}{Moreover, in view of a molecular trimer, applications
in terms of targeted energy transfer (introduced in \cite{Kopidakis:2001p58,Tsironis:2003p59})
with a threshold are conceivable, e.g.~in the field of biophysics
or biomolecular engineering.}

\subsubsection{\textcolor{black}{Conclusions}}

\textcolor{black}{The threshold and the tunneling process during the
collision of a DB with a lattice excitation (e.g.~a moving breather)
was described analytically by defining the $2$-dimensional Peierls-Nabarro
energy landscape. The PN landscape restricts the dynamics of the trimer
and the accessible region of the phase space. This restriction of
the dynamics becomes very pronounced at the destabilization threshold,
which is identified with a rim in the PN landscape. The effect is
found for a broad range of the phase difference $\delta\phi_{12}$
between the colliding objects.}
\begin{acknowledgments}
The authors thank Boston University, where the bulk of this work was
done, for hospitality and support. One of us (DKC) thanks the Max
Planck Institute for Dynamics and Self-Organization for hospitality
during the initial stages of the project and the Aspen Center for
Physics for hospitality during the completion of the work. We gratefully
acknowledge useful conversations with Theo Geisel, Ragnar Fleischmann
and Cristiane de Morais Smith.\bibliographystyle{apsrev}
\bibliography{/Users/holgerh/mpi/bib/papers_jabref}

\end{acknowledgments}

\end{document}